\documentclass{aa}

\usepackage{graphicx}
\usepackage{txfonts}
\usepackage{hyperref}
\bibpunct{(}{)}{;}{a}{}{,}

\usepackage{amsmath}	
\usepackage{amssymb}	
\usepackage{multicol}        
\usepackage{bm}		
\usepackage{pdflscape}	

\usepackage{comment}

\usepackage[T1]{fontenc}

\newcommand{\ngc}[1]{NGC~#1}

\newcommand{\srca}{NGC~4388}
\newcommand{\srcb}{NGC~2110}

\newcommand{\xmm}{{\em XMM--Newton}}
\newcommand{\nus}{{\em NuSTAR}}
\newcommand{\chandra}{{\em Chandra}}
\newcommand{\suz}{{\em Suzaku}}
\newcommand{\integral}{{\em INTEGRAL}}
\newcommand{\sax}{{\em BeppoSAX}}
\newcommand{\swift}{{\em Swift}}

\newcommand{\aer}[3]{$#1^{+ #2}_{- #3}$}
\newcommand{\aerm}[3]{#1^{+ #2}_{- #3}}
\newcommand{\aerexp}[4]{$#1^{+ #2}_{- #3} \times 10^{#4}$}
\newcommand{\ser}[2]{$#1 \pm #2$}
\newcommand{\serm}[2]{#1 \pm #2}
\newcommand{\serexp}[3]{($#1 \pm #2) \times 10^{#3}$}

\newcommand{\linf}[1]{$> #1$}
\newcommand{\expo}[2]{$ #1 \times 10^{#2}$}
\newcommand{\expom}[2]{#1 \times 10^{#2}}
\newcommand{\tento}[1]{$10^{#1}$}
\newcommand{\tentom}[1]{10^{#1}}

\newcommand{\chisq}{\chi^{2}}
\newcommand{\rchisq}{\chi^{2}/\textrm{dof}}
\newcommand{\dchi}{\Delta \chi^{2}}
\newcommand{\ddof}{\Delta \textrm{dof}}

\newcommand{\cut}{E_{\textrm{c}}}
\newcommand{\nh}{N_{\textrm{H}}}
\newcommand{\cf}{\textrm{C}_{\textrm{F}}}

\newcommand{\msun}{M$_{\odot}$}

\newcommand{\kalfa}{K~$\alpha$}

\newcommand{\fek}{Fe~K$\alpha$}

\newcommand{\ione}[2]{#1~{\sc #2}}

\newcommand{\xspec}{{\sc xspec}}

\newcommand{\compps}{{\sc compps}}

\newcommand{\pexrav}{{\sc pexrav}}

\newcommand{\kte}{kT_{\textrm{e}}}
\newcommand{\tetae}{\Theta_{\textrm{e}}}

\newcommand{\elmass}{m_{\textrm{e}}}

\newcommand{\lumcgs}{ergs~s$^{-1}$}

\newcommand{\sqcm}{cm$^{-2}$}

\newcommand{\refl}{\mathcal{R}}

\begin{document} 

\title{The coronal temperature of NGC~4388 and NGC~2110 measured with \textit{INTEGRAL}\thanks{The reduced spectra (FITS files) are available at the CDS via anonymous ftp to cdsarc.u-strasbg.fr (130.79.128.5) or via http://cdsweb.u-strasbg.fr/cgi-bin/qcat?J/A+A/...}}

\titlerunning{The coronal temperature of NGC~4388 and NGC~2110}
\authorrunning{F. Ursini et al.}


   \author{F. Ursini\inst{1},
           L. Bassani\inst{1},
           A. Malizia\inst{1},
           A. Bazzano\inst{2},
           A.~J. Bird\inst{3},
           J.~B. Stephen\inst{1},
		\and
          P. Ubertini\inst{2}
          }

   \institute{
   	INAF-Osservatorio di astrofisica e scienza dello spazio di Bologna, Via Piero Gobetti 93/3, 40129 Bologna, Italy\\
    \email{francesco.ursini@inaf.it}
    \and
    INAF-Istituto di Astrofisica e Planetologia Spaziali, via Fosso del Cavaliere, 00133 Roma, Italy
    \and
    School of Physics and Astronomy, University of Southampton, SO17 1BJ, UK
             }

   \date{Received ...; accepted ...}

 
  \abstract
   {}
   {
   	We aim to measure the physical properties of the hot X-ray corona of two active galactic nuclei, \ngc{4388} and \ngc{2110}.
   	}
   {We analysed the hard X-ray (20--300 keV) \integral\ spectrum in conjunction with archival \xmm\ and \nus\ data.}
   {The X-ray spectrum of both sources is phenomenologically well described by an absorbed cut-off power law. In agreement with previous results, we find no evidence of a Compton reflection component in these sources. We obtain a high-energy cut-off of \aer{200}{75}{40} keV for \ngc{4388} and \aer{320}{100}{60} keV for \ngc{2110}. A fit with a thermal Comptonisation model yields a coronal temperature of \aer{80}{40}{20} keV and \aer{75}{20}{15} keV, respectively, and an optical depth of approximately two, assuming a spherical geometry. The coronal temperature and luminosity of both sources are consistent with pair production that acts as a thermostat for the thermal plasma. These results emphasise the importance of good signal-to-noise X-ray data above 100 keV to probe the high-energy emission of AGNs. }
   {}


   \maketitle
%

\section{Introduction}
One of the main components of active galactic nuclei (AGNs) is a hot corona, which is thought to produce the X-ray emission via thermal Comptonisation of optical and UV photons from the accretion disc. This process naturally explains the observed power-law shape of the X-ray spectra of AGNs and their high-energy cut-off, which is related to the coronal temperature. The cut-off has often been observed at $\sim 100$ keV 
thanks to both past and current X-ray missions, such as the \textit{Compton Gamma-Ray Observatory}'s Oriented Scintillation Spectrometer Experiment \cite[OSSE;][]{zdziarski2000}, \sax\ \cite{perola2002}, \swift's Burst Alert Telescope \cite[BAT;][]{bat70,ricci2017}, \integral\ \citep{malizia2014}, and \nus\ \citep[][and references therein]{fabian2015,tortosa2018}.  
Good constraints on the high-energy cut-off have been, and are still, obtained for a number of AGNs with \nus. This leads to measurements of the coronal temperature to range from 15--20 keV in the `coolest' sources \citep{kara2017,tortosa2017,buisson2018,turner2018} to $\sim 400$ keV in NGC 5506 \citep{matt20155506}. However, in many cases only lower limits to the cut-off have been reported \citep[e.g.][]{kamraj2018} that can still be useful to constrain the geometrical and physical parameters of the hot corona \cite[e.g.][]{matt2014ark120,marinucci2015,7213}.

Thanks to its high-energy coverage, the Imager on-board \integral\ \cite[IBIS;][]{ibis} allows us to obtain good measurements of the cut-off energy \citep{panessa2008,molina2009,derosa2012,malizia2014} and of the coronal temperature \citep{lubinski2016}. 
In this paper, we present the \integral's IBIS spectrum together with archival \xmm\ and \nus\ observations of two of the brightest Seyfert galaxies in the hard X-rays: \ngc{4388} and \ngc{2110}. In Sect. \ref{sec:data} we report the data selection and reduction. In Sect. \ref{sec:analysis} we report the spectral analysis. In Sect. \ref{sec:discussion} we discuss the results and 
summarise the conclusions. The main properties of the two sources are summarised below.

\subsection{NGC 4388}
\ngc{4388} is a nearby \cite[$z=0.00842$,][]{lu1993}, X-ray bright Seyfert galaxy, hosting a supermassive black hole of \serexp{8.5}{0.2}{6} \msun\ \cite[measured from the water maser; see][]{kuo2011}. 
This source has been observed by 
all of the major X-ray satellites. The hard X-ray spectrum is moderately absorbed by a column density of a few $\times \tentom{23}$ \sqcm, which is found to be variable on short time-scales \citep{elvis2004}. The soft X-ray spectrum below $\sim 2-3$ keV is dominated by emission of a hot plasma in an extended X-ray nebula at kpc scales \citep{matt1994,iwasawa2003,beckmann2004,bianchi2006}. No Compton reflection hump has been detected with \nus\ \citep{kamraj2017} despite the presence of a strong \fek\ emission line, that likely originates from Compton-thin material \citep{kamraj2017}. Past results indicate the presence of a high-energy cut-off at $\sim 200$ keV, albeit with some uncertainties (see Table \ref{tab:past}). From \integral\ data taken in 2003 in combination with \xmm\ spectra, \cite{beckmann2004} reported a lower limit of 180 keV to the high-energy cut-off. The same result is obtained by \cite{derosa2012}, also from \integral+\xmm\ data. Using high-energy \integral\ and BAT data, \cite{molina2013} report a high-energy cut-off of \aer{202}{111}{54} keV. 
From \integral\ data taken from 2003 to 2009, 
\cite{fedorova2011} report strong variations in both the flux and spectral slope in the 20--60 keV band on a few months time-scale. These authors also report a high-energy cut-off at $\sim 200$ keV with some indications of variability between 80--100 keV and \linf{320} keV, 
but with limited statistics. From \integral\ data taken from 2003 to 2010, \cite{lubinski2016} report a temperature of \aer{53}{17}{9} keV and an optical depth of \aer{2.7}{0.6}{0.9}. However, no constraint on the cut-off has been found with \nus\ \citep{kamraj2017} and only a lower limit of $\sim 100$ keV was reported from \sax\ \citep{risaliti2002} and \swift's BAT \citep{ricci2017}.
From the analysis of the 58-month \swift's BAT light curves, \cite{caballero2012} report the detection of hard X-ray spectral variability in the 14--195 keV band. 
However, 
\cite{soldi2014} do not find strong evidence of long-term hard X-ray spectral variability from BAT data up to 66 months.

\subsection{NGC 2110}
\ngc{2110} is another nearby \cite[$z=0.00779$,][]{gallimore1999}, X-ray bright Seyfert galaxy.
\cite{diniz2015} report a black hole mass of \aerexp{2.7}{3.5}{2.1}{8} \msun, from the relation with the stellar velocity dispersion.  
From \sax\ data, \cite{malaguti1999} found the X-ray spectrum to be affected by complex absorption. This has been later confirmed 
by \cite{evans2007}, who find the \chandra+\xmm\ data to be well fitted with a neutral, three-zone, partial-covering absorber. \cite{rivers2014} find the \suz\ data to be well fitted with a stable full-covering absorber plus a variable partial-covering absorber. A soft excess below 1.5 keV is also present \citep{evans2007}, and possibly due to extended circumnuclear emission seen with \chandra\ \citep{evans2006}. No Compton reflection hump has been detected with \suz\ \citep{rivers2014} or \nus\ \citep{marinucci2015}, despite the presence of a complex \fek\ line. According to the multi-epoch analysis of \cite{marinucci2015}, the \fek\ line is likely the sum of a constant component (from distant, Compton-thick material) and a variable one (from Compton-thin material). Concerning the high-energy cut-off, ambiguous results have been reported in literature (see Table \ref{tab:past}). \cite{ricci2017} report a value of \aer{448}{63}{55} keV, while \cite{lubinski2016} report a coronal temperature of \aer{230}{51}{57} keV and an optical depth of \aer{0.52}{0.14}{0.13}. From 2008-2009 \integral\ data, \cite{beckmann2010} report a cut-off of $\sim 80$ keV with a hard photon index, but these results are not confirmed by \nus\ \citep{marinucci2015}. Indeed, only lower limits to the high-energy cut-off have been found with \nus\ \citep[210 keV:][]{marinucci2015}, \suz\ \citep[250 keV:][]{rivers2014} and \sax\ \citep[143 keV:][]{risaliti2002}. 
No hard X-ray spectral variability has been detected by \cite{caballero2012} and \cite{soldi2014} from BAT data, despite the significant flux variability.

 \begin{table}
 	\caption{Previous constraints on high-energy cut-off and coronal temperature (both in keV) reported in literature.}              
 	\label{tab:past}     
 	\centering                                      
 	\begin{tabular}{c c c }          
 		\hline\hline   
 		&\ngc{4388} & \ngc{2110} \\
 		\hline     
 		$\cut$  & \linf{145}\tablefootmark{a} & \linf{143}\tablefootmark{a} \\
				&\linf{180}\tablefootmark{b}&\ser{82}{9}\tablefootmark{g} \\ 		
 		 		&\aer{209}{44}{32}\tablefootmark{c} &\linf{250}\tablefootmark{h}\\[0.5ex]
 		 		&\aer{202}{111}{54}\tablefootmark{d}&  \linf{210}\tablefootmark{i}  \\
		 		&\linf{104}\tablefootmark{e}&\aer{448}{63}{55}\tablefootmark{e}\\[0.5ex]
 		\hline
 		$\kte$ & \aer{53}{17}{9}\tablefootmark{f}& \aer{230}{51}{57}\tablefootmark{f}\\
 		&&\ser{190}{130}\tablefootmark{i}\\
 		\hline
 	\end{tabular}
 	\tablefoot{
	 	\tablefoottext{a}{\cite{risaliti2002}}
	 	\tablefoottext{b}{\cite{beckmann2004}, \cite{derosa2012} }
	 	\tablefoottext{c}{\cite{fedorova2011}}
	 	\tablefoottext{d}{\cite{molina2013}}
	 	\tablefoottext{e}{\cite{ricci2017}}
		\tablefoottext{f}{\cite{lubinski2016}}
		\tablefoottext{g}{\cite{beckmann2010}}
		\tablefoottext{h}{\cite{rivers2014}}
		\tablefoottext{i}{\cite{marinucci2015}}
 	}
 \end{table}

\section{Data selection and reduction}\label{sec:data}
For both sources, we collected the archival \xmm\ and \nus\ data to complement the IBIS spectrum (see Table \ref{tab:log}).
The \xmm\ data were processed using the \xmm\ Science Analysis System (\textsc{sas} v18). We focused on the EPIC-pn data because of the much larger effective area compared with the MOS detectors. The source extraction radii and screening for high-background intervals were determined through an iterative process that maximises the signal-to-noise ratio \cite[see][]{pico2004}. The background were extracted from circular regions with a radius of 50 arcsec, while the source extraction radii were in the range 20--40 arcsec. The spectra were binned to have at least 30 counts per spectral bin and not oversampling the instrumental resolution by a factor larger than 3.

The \nus\ data were reduced using the standard pipeline (\textsc{nupipeline}) in the \nus\ Data Analysis Software (\textsc{nustardas}, v1.8.0), using calibration files from \nus\ {\sc caldb} v20190410. We extracted the spectra 
using the standard tool {\sc nuproducts} for each of the two hard X-ray detectors, which reside in the corresponding focal plane modules A and B (FPMA and FPMB). We extracted the source data from circular regions with a radius of 75 arcsec, and the background from a blank area close to the source. The spectra were binned to have a signal-to-noise ratio larger than 3 in each spectral channel and not to oversample the instrumental resolution by a factor greater than 2.5. The spectra from the two detectors were analysed jointly but not combined.

Both \ngc{4388} and \ngc{2110} are detected with a good signal-to-noise in the hard X-ray band with IBIS ($> 6 \sigma$ in the 150--300 keV band, see Table \ref{tab:src}). 
Here we use the data collected by the \integral\ Soft Gamma-Ray Imager \citep[ISGRI:][]{isgri}, that is the low-energy camera of the IBIS telescope. We used data up to \integral\ orbit 1500, that is from the launch in 2002 to the end of January 2015. ISGRI images for each available pointing were generated in 14 energy bands  using the ISDC offline scientific analysis (OSA) software \citep{osa}, version 10.2. 
Count rates at the position of the source were extracted from individual images to provide light curves in 14 energy bands between 20 and 300 keV. Weighted mean fluxes were then extracted in each band and combined to produce an average source spectrum \cite[see][for details]{bird2007,bird2010}.
Corresponding weighted response (arf) files for each source were created by weighting the published matrices for each response validity period according to the photon fluence from the source during that validity period. This method takes into account both the different exposure in each validity period, and any source variability present. A single rmf response is used, corresponding to the standard rmf, rebinned according to the 14 channels used for spectral analysis.
For \ngc{4388}, the spectrum was derived from 4770 pointings with a total on-source time (not adjusted for off-axis response) of 7.5 Ms. The corresponding figures for \ngc{2110} are 2249 pointings for 3.8 Ms total on-source time. In both cases, the observations are spread reasonably uniformly across the time period analysed.  
 
 \begin{table}
 	\caption{IBIS source significance of \ngc{4388} and \ngc{2110} in three energy bands: 100--120 keV, 120--150 keV, 150--300 keV.}              
 	\label{tab:src}     
 	\centering                                      
 	\begin{tabular}{c c c c}          
 		\hline\hline   
 		&$\sigma_{100-120}$&$\sigma_{120-150}$&$\sigma_{150-300}$ \\
 		\hline     
 		\ngc{4388}  &22.2&17.6&8.6\\
 		\ngc{2110}  &17.6&12.9&6.6\\
 		\hline
 		
 	\end{tabular}
 \end{table}

\begin{table*}
	\caption{Low-energy X-ray data of \srca\ and \srcb\ analysed in this work.}              
	\label{tab:log}     
	\centering                                      
	\begin{tabular}{c c c c c}          
		\hline\hline    
		Satellite & Obs. Id. & Start time (\textsc{utc}) & Net exp. & label  \\  
		&&yyyy-mm-dd&(ks) &\\  
		\hline          
				\multicolumn{5}{c}{NGC 4388}   \\                          
		\xmm & 0110930301 & 2002-07-07 & 3.8 & XMM1\\      
		\xmm & 0110930701 & 2002-12-12 & 7.3 & XMM2\\
		\xmm & 0675140101 & 2011-06-18 & 37.7& XMM3 \\
		\nus & 60061228002 & 2013-12-27 &  21.4 & NUS  \\
		\hline      
		\multicolumn{5}{c}{NGC 2110}  \\
		\xmm & 0145670101 & 2003-03-05 & 44.5 & XMM \\
		\nus & 60061061002 & 2012-10-05 & 15.5& NUS1 \\
		\nus & 60061061004 & 2013-02-14 & 12.0& NUS2 \\
		\hline
		                                   
	\end{tabular}
\end{table*}

\section{Spectral analysis}\label{sec:analysis}
Spectral analysis and model fitting was carried out with the \xspec\ 12.10 package \citep{arnaud1996}, using the $\chisq$ minimisation technique. Errors are quoted at the 90 per cent confidence level for one interesting parameter. We assumed the element abundances of \cite{angr} and the photoelectric absorption cross-sections of \cite{vern}. We fitted the IBIS spectra over the full 20--300 keV energy band. 
\subsection{NGC 4388}
The emission of this source below $\sim 3$ keV is known to be dominated by emission from a hot plasma in an extended X-ray nebula at kpc scales \citep{beckmann2004}. Therefore, since our main focus is the AGN-dominated hard X-ray emission, we restricted the broad-band spectral analysis to the data above 4 keV. We fitted the \xmm\ and \nus\ spectra over the 4--10 keV and 4--79 keV energy bands, respectively.
\subsubsection{The IBIS spectrum}
As a first step, we focused on the IBIS spectrum, with a twofold purpose: first, testing for the presence of a curvature at high energies and, secondly, a consistency check with the low-energy spectra.
We performed fits assuming absorption by a column density $\nh= \expom{2.7}{23}$ \sqcm\ \citep{beckmann2004} since this value is not constrained by the data above 20 keV. 

\begin{figure} 
	\includegraphics[width=\columnwidth]{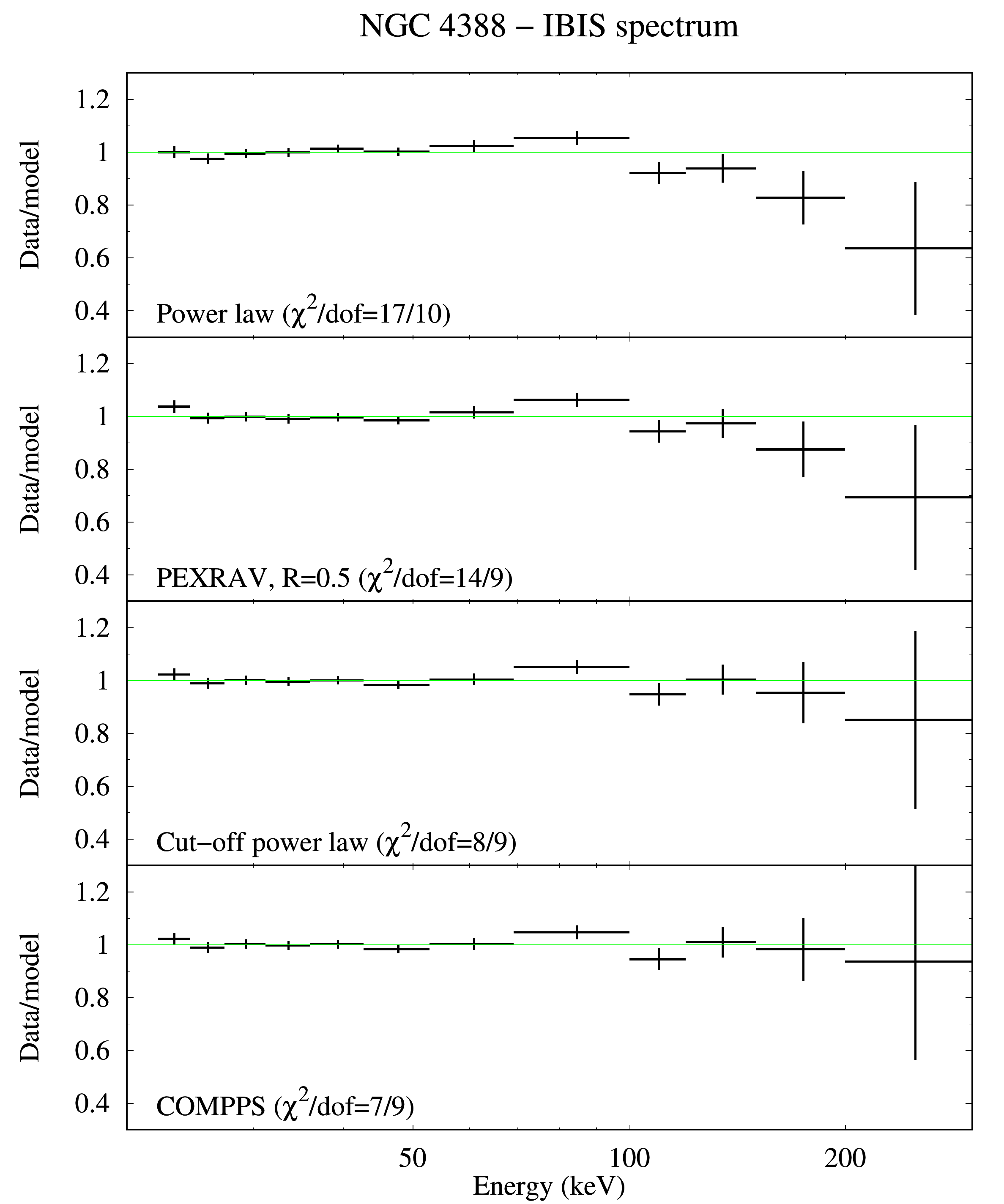}
	\caption{\label{fig:ratios} Residuals of fits of IBIS spectrum with different models. Upper panel: simple power law. Second panel: power law plus reflection (\pexrav). Third panel: exponentially cut-off power law. Lower panel: thermal Comptonisation model (\compps). }
\end{figure}

\begin{figure} 
	\includegraphics[width=\columnwidth]{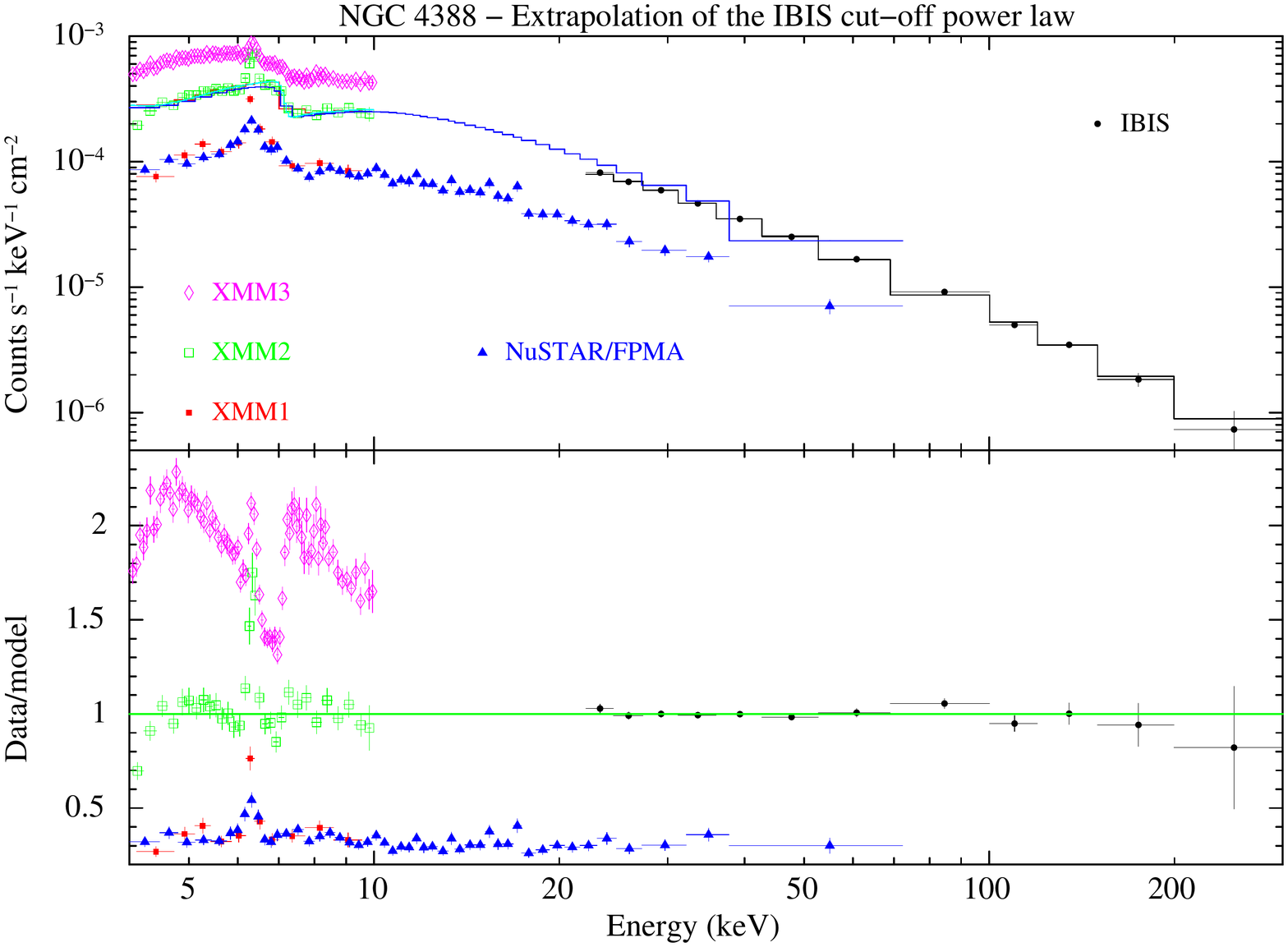}
	\caption{\label{fig:extrap_4388} Upper panel: \xmm\ and \nus\ spectra of NGC 4388 with the cut-off power law that best fits IBIS. Lower panel: data/model ratio. Only \nus's FPMA data are shown for clarity. The data were binned for plotting purposes.}
\end{figure}

Fitting the IBIS data with a power law, we obtain a poor fit ($\rchisq=17/10$) with negative residuals above 100 keV (Fig. \ref{fig:ratios}, first panel). To reproduce this curvature, we first tried to include Compton reflection, replacing the cut-off power law with \pexrav\ \cite[]{pexrav}. This model includes Compton reflection off a neutral slab of infinite column density. We fixed the inclination angle at 60 deg since the fit was not sensitive to this parameter. We first fixed the reflection fraction ($\mathcal{R} = \Omega / 2 \pi$, where $\Omega$ is the solid angle subtended by the reflector) at 0.5. The model also includes an exponential high-energy cut-off, which we left free to vary. This yields a slightly better fit ($\rchisq=14/9$), but still with the indication of a curvature (Fig. \ref{fig:ratios}, second panel). Indeed, leaving $\mathcal{R}$ free to vary, the value is pegged at zero with an upper limit of 0.3. Then, we replaced \pexrav\ with a cut-off power law (equivalent to \pexrav\ with $\mathcal{R}=0$) finding a good fit ($\rchisq=8/9$) and a cut-off energy of \aer{300}{400}{100} keV. Finally, we replaced the phenomenological cut-off power law with the Comptonisation model \compps\ \citep{compps}. We obtain a good fit ($\rchisq=7/9$) and a temperature of \aer{100}{150}{40} keV. 
We also repeated the same analysis assuming the partial-covering absorption with $\nh= \expom{6.5}{23}$ \sqcm\ and a covering factor of 0.9, consistent with \nus\ data \citep{kamraj2017}, obtaining analogous results. 
Then we extrapolated the best-fitting cut-off power law down to 4 keV, without re-fitting, to compare it with the \xmm\ and \nus\ spectra (Fig. \ref{fig:extrap_4388}). Despite the flux variability, the XMM1, XMM2, and NUS spectral shape appears in good agreement with the extrapolation, with minor discrepancies likely due to an imperfect modelling of absorption. XMM3 instead shows a significant difference in spectral shape. 

\subsubsection{The broad-band fit}
As a second step, we performed a broad-band fit including the lower energy data from \xmm\ and \nus. We used the XMM1, XMM2, and NUS spectra, that is to say those in good agreement with IBIS. XMM3, despite having the deepest exposure, likely represents a different spectral state compared with the average and is discussed in Appendix \ref{appendix}. 

\begin{figure} 
	\includegraphics[width=\columnwidth]{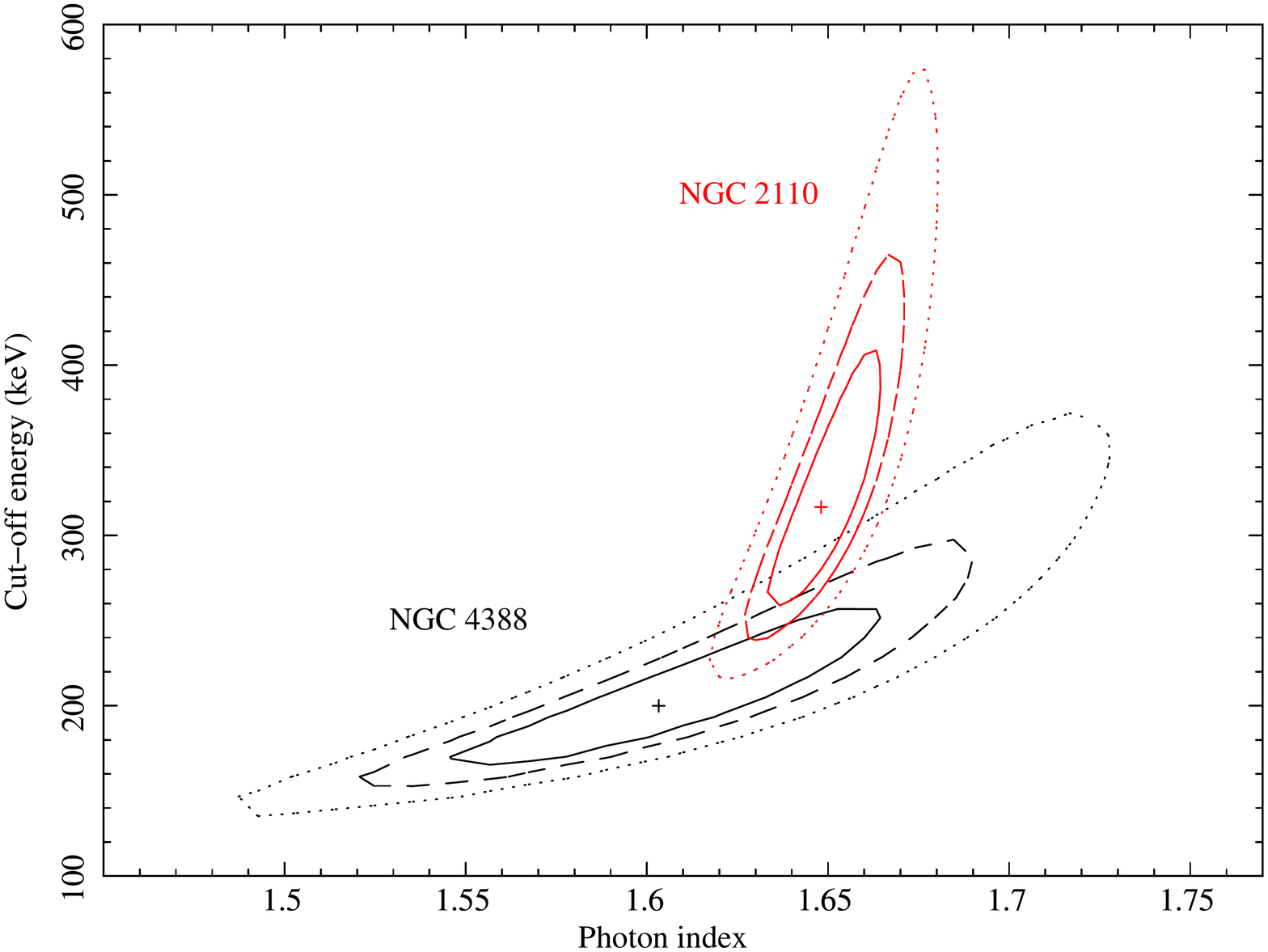}
	\caption{\label{fig:cont_cutoffpl} Contour plots of cut-off energy vs. photon index for NGC 4388 (black) and NGC 2110 (red). Solid, dashed, and dotted lines correspond to 68, 90, and 99 per cent confidence level, respectively.}
\end{figure}

First, we fitted the data with a phenomenological model. Following \cite{kamraj2017}, we used a model consisting of a cut-off power law plus a Gaussian \fek\ line, modified by partial covering absorption. The model reads \textsc{pcfabs(cutoffpl+zgauss)} in \xspec\ terminology. We included a cross-normalisation constant free to vary among the XMM1, XMM2, and NUS observations, to account for the flux variations. We first assumed all the other parameters to be constant, meaning that they were tied among the different spectra. We obtained a good fit ($\rchisq=455/435$), but with some residuals near 6.4 keV indicating variability of the iron line flux not related to the primary continuum. We thus left the normalisation of the Gaussian line free to vary among the different spectra. We obtained an excellent fit ($\rchisq=433/432$), with $\Gamma \simeq 1.6$ and a well constrained cut-off energy of \aer{200}{75}{40} keV. 
We found no significant improvement by leaving the column density $\nh$, the covering factor $\cf$, or the photon index free to vary among the different observations. 
The best-fitting parameters are reported in Table \ref{tab:fits_4388}, while the contour plots of the cut-off energy and photon index are shown in Fig. \ref{fig:cont_cutoffpl}. 
We also tried to include a reflection component, replacing the cut-off power law with \pexrav\ and leaving the reflection fraction $\refl$ free and tied among the observations. The fit is not improved 
and we obtain only an upper limit $\refl < 0.12$. We note that the large equivalent width of the \fek\ line is consistent with that reported by \cite{kamraj2017}, who suggest the presence of a large amount of Compton-thin line-emitting material.

Then, we fitted the data using a more physical Comptonisation model. We replaced the cut-off power law with \compps, assuming a spherical geometry ($\textsc{geom}=0$). We assumed a seed photon temperature of 100 eV. Fixing this parameter at other values, like 10 eV, does not alter the results significantly. We fitted for the electron temperature $\kte$ and the Compton parameter $y = 4\tau (\kte/m_{\textrm{e}} c^2)$. 
We chose to use the Compton parameter instead of the optical depth to minimise the model degeneracy since the temperature and optical depth are generally correlated in the fitting procedure \citep[e.g.][]{poptestingcompt2001,pop2013mrk509}.

\begin{figure} 
	\includegraphics[width=\columnwidth]{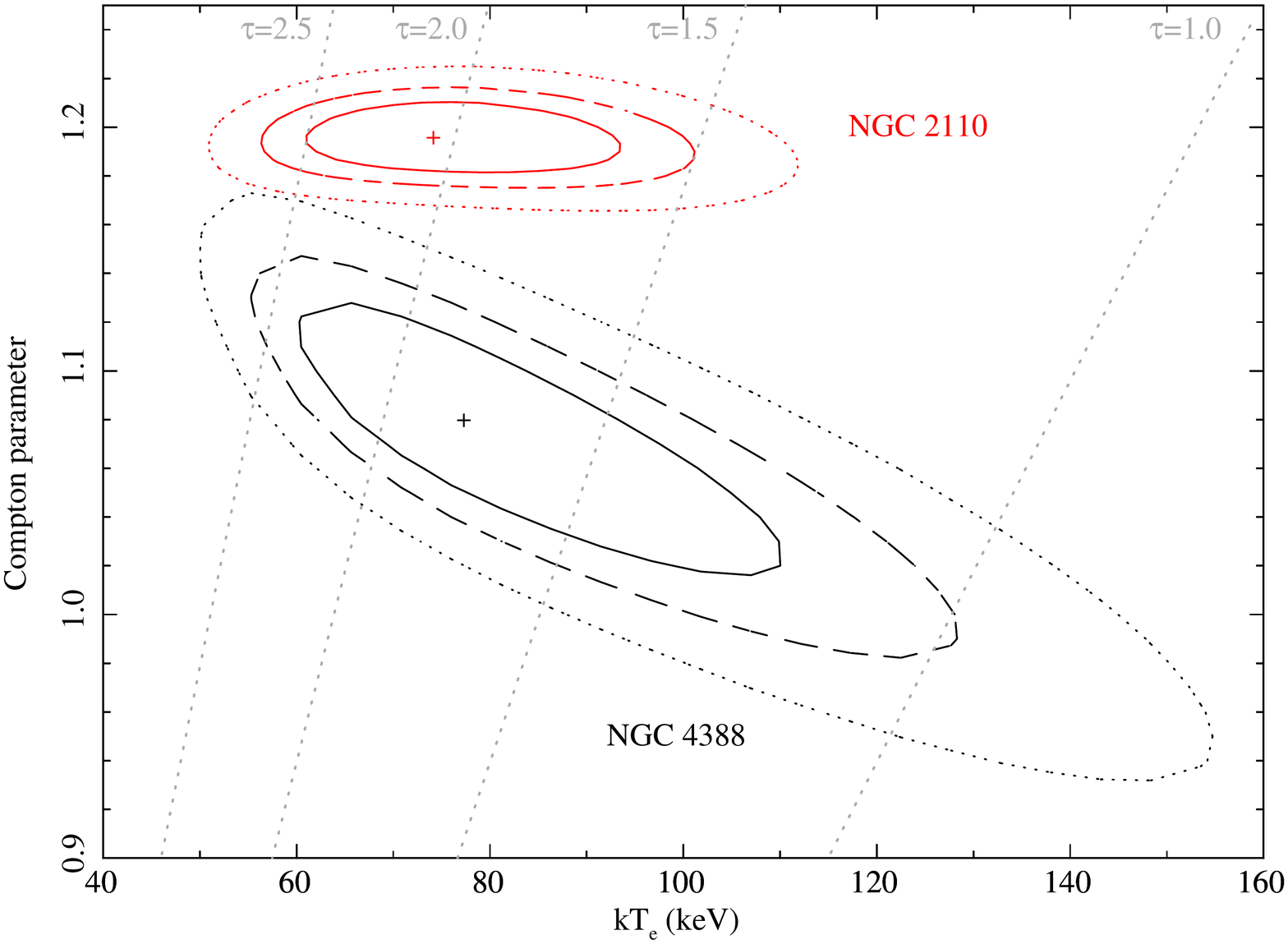}
	\caption{\label{fig:cont_compps} Contour plots of Compton parameter $y$ vs. electron temperature $\kte$ for NGC 4388 (black) and NGC 2110 (red). Solid, dashed, and dotted lines correspond to 68, 90, and 99 per cent confidence level, respectively. Grey dotted lines correspond to constant values of optical depth.}
\end{figure}

We obtained a very good fit ($\rchisq=444/432$), with $\kte = \aerm{80}{40}{20}$ keV and $y = \aerm{1.08}{0.05}{0.08}$, corresponding to an optical depth $\tau=\serm{1.7}{0.7}$. The best-fitting parameters ($\kte, y$, and normalisation) are reported in Table \ref{tab:fits_4388}, while the $\kte - y$ contour plots are shown in Fig. \ref{fig:cont_compps}. The cross-normalisations, absorption and iron line parameters are consistent within the errors with the cut-off power law fit.


\begin{figure} 
	\includegraphics[width=\columnwidth]{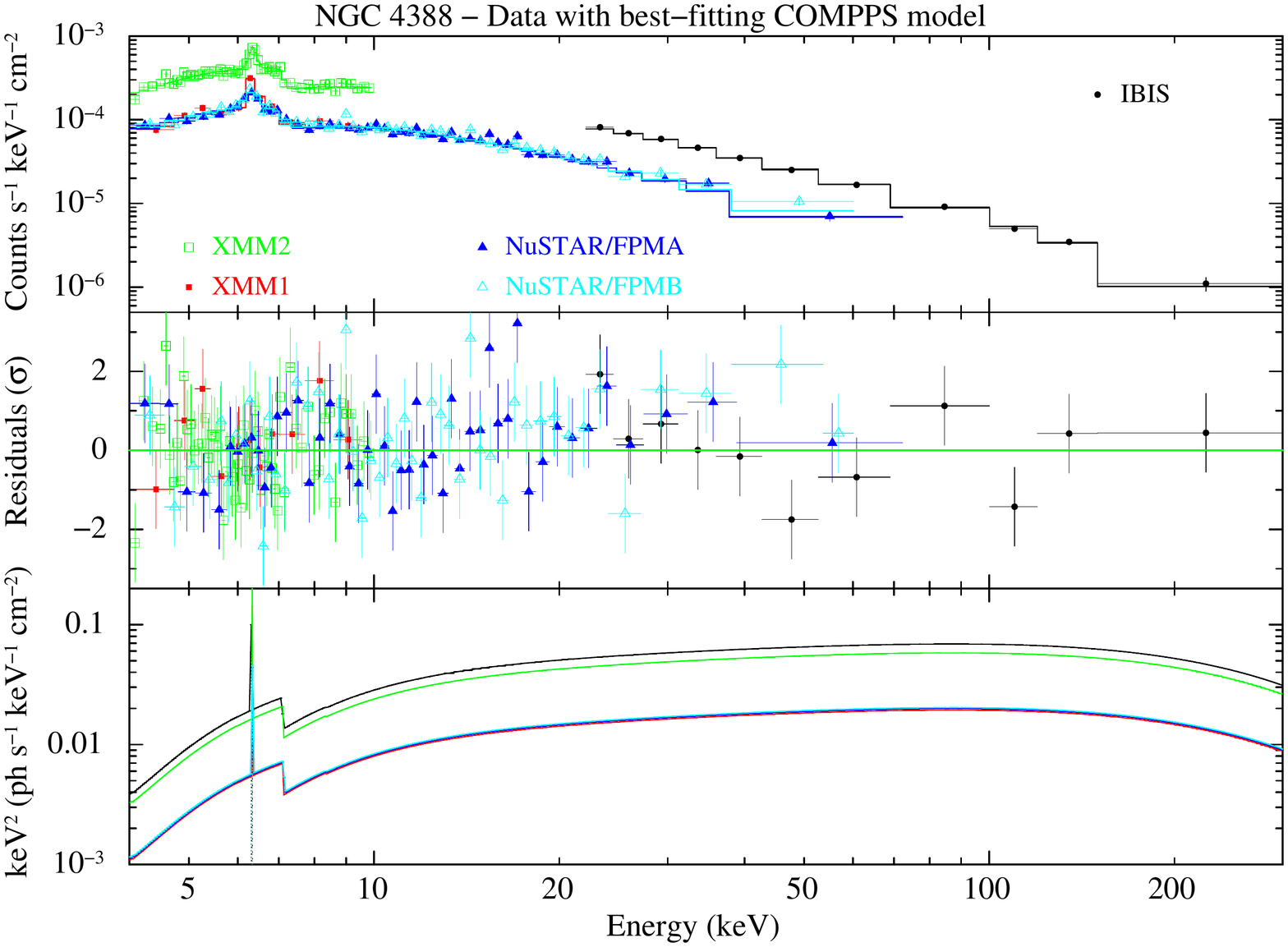}
	\caption{\label{fig:plot_4388} Upper panel: spectra of NGC 4388 with best-fitting \compps\ model. Second panel: residuals, plotted as $\Delta \chi = $ (data-model)/error. Third panel: best-fitting model $E^2 f(E)$. The data were binned for plotting purposes.}
\end{figure}

\begin{table*}
	\begin{center}
		\caption{ \label{tab:fits_4388} Best-fitting parameters for NGC 4388. (t) denotes a parameters tied between the spectra.} 
		\begin{tabular}{ l c c c c} 
			\hline  \hline 
			 & all obs. & XMM1 & XMM2 & NUS \\
			\hline
			\multicolumn{5}{c}{\textsc{cutoffpl} model }\\
			\hline     
			\noalign{\smallskip}
			$\Gamma$ & \aer{1.60}{0.07}{0.06}&(t)&(t)&(t) \\[0.5ex]
			$\cut$ (keV)& \aer{200}{75}{40}&(t)&(t)&(t) \\[0.5ex]
			$N_{\textsc{pow}}$ (\tento{-2}) & \aer{1.8}{0.4}{0.2}&(t)&(t)&(t)  \\     
			\noalign{\smallskip}
			$K_{\textrm{IBIS--pn}}$  && \ser{0.29}{0.02} &\ser{0.86}{0.04}& -       \\
			$K_{\textrm{IBIS--NusA}}$  &&-  &-&   \ser{0.29}{0.01}        \\  
			$K_{\textrm{IBIS--NusB}}$  &&-  &-&   \ser{0.30}{0.01}        \\   
			\noalign{\smallskip}
			$\nh$ (\tento{23} \sqcm) & \aer{3.8}{0.5}{0.2}&(t)&(t)&(t)\\[0.5ex]        
			$\cf$   & \aer{0.95}{0.03}{0.02} &(t)&(t)&(t) \\  
			\noalign{\smallskip}
			$E_{\textsc{ga}}$ (keV) & \aer{6.39}{0.02}{0.01}&(t)&(t)&(t) \\
			$N_{\textsc{ga}}$ (\tento{-4}) && \ser{3.5}{0.8}  & \aer{1.6}{0.3}{0.2} & \ser{2.7}{0.6} \\
			EW$_{\textsc{ga}}$ (eV) && \aer{390}{100}{90}  & \ser{180}{30} & \ser{300}{70} \\
			\noalign{\smallskip}
			$\rchisq$ &433/432&&&\\
			\hline  
			\multicolumn{5}{c}{\textsc{compps} model }\\
			\hline   
			\noalign{\smallskip}
			$\kte$ (keV)& \aer{80}{40}{20}&(t)&(t)&(t) \\[0.5ex]
			$y$ & \aer{1.08}{0.05}{0.08}&(t)&(t)&(t) \\[0.5ex]
			$N_{\textsc{compps}}$ (\tento{4})  &\aer{6.5}{1.5}{1.0} &(t)&(t)&(t) \\  
			\noalign{\smallskip} 
			$\rchisq$ &444/432&&&\\    
			\hline   
		\end{tabular}
	\end{center}
\end{table*}

\subsection{NGC 2110}
The soft X-ray emission of \ngc{2110} below $\sim 2-3$ keV is affected by variable absorption from a complex medium \citep{evans2007,rivers2014}. Therefore, we restricted the spectral analysis to the data above 3 keV. We fitted the \xmm\ and \nus\ spectra over the 3--10 keV and 3--79 keV energy bands.
\subsubsection{The IBIS spectrum}
As we did for \ngc{4388}, we first focused on the IBIS spectrum.
The spectrum above 20 keV is not strongly altered by absorption, however we assumed $\nh= \expom{4}{22}$ \sqcm\ \citep{marinucci2015}. 

\begin{figure} 
	\includegraphics[width=\columnwidth]{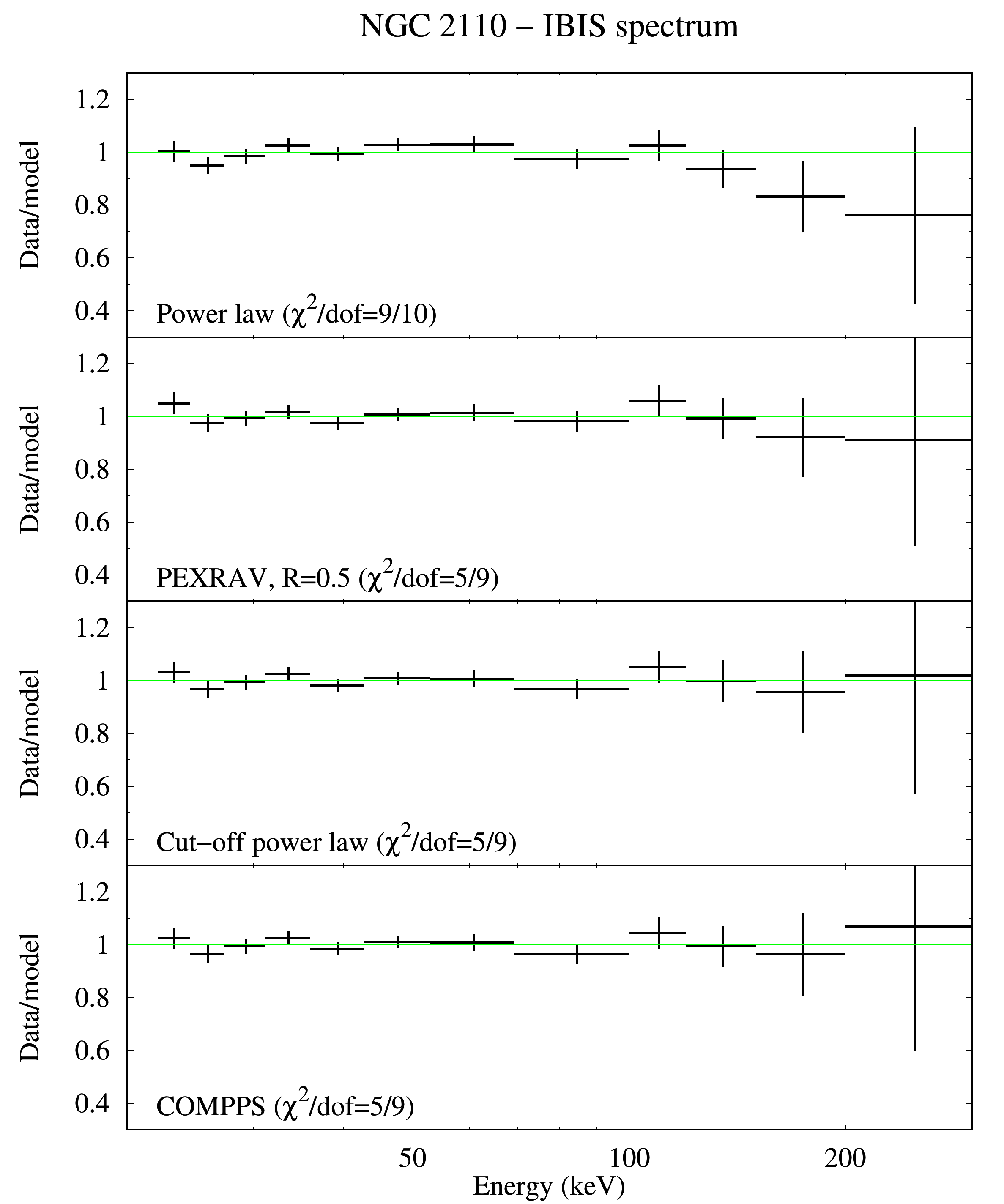}
	\caption{\label{fig:ratios2110} Residuals of fits of IBIS spectrum with different models. Upper panel: simple power law. Second panel: power law plus reflection (\pexrav). Third panel: exponentially cut-off power law. Lower panel: thermal Comptonisation model (\compps). }
\end{figure}

\begin{figure} 
	\includegraphics[width=\columnwidth]{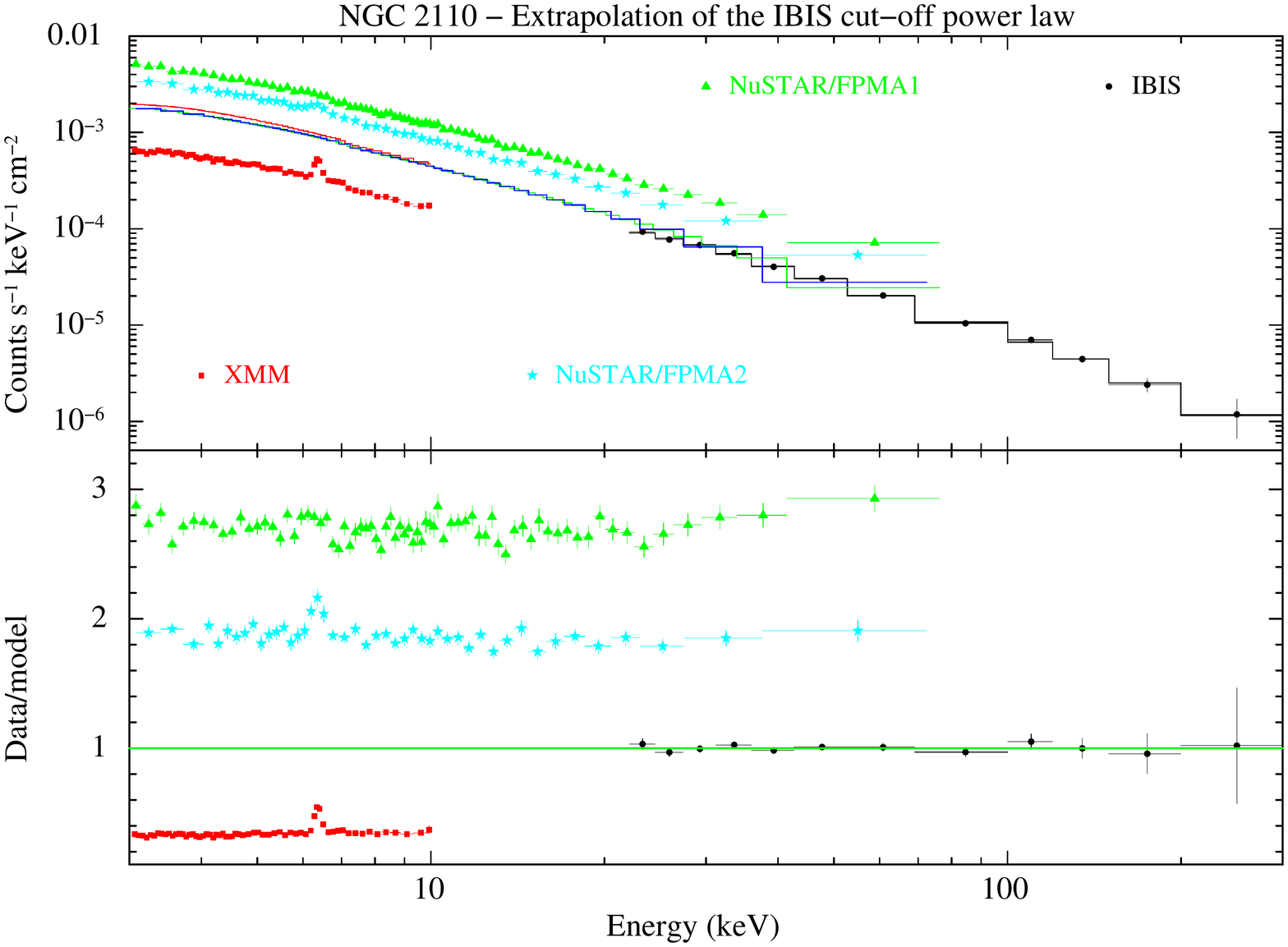}
	\caption{\label{fig:extrap_2110} Upper panel: \xmm\ and \nus\ spectra of NGC 2110 with the cut-off power law that best fits IBIS. Lower panel: data/model ratio. Only \nus's FPMA data are shown for clarity. The data were binned for plotting purposes.}
\end{figure}

Fitting the data with a power law, we obtained a statistically good fit ($\rchisq=9/10$) with a hint of a curvature above 120 keV (Fig. \ref{fig:ratios2110}, first panel).
We also tested three other models, as done for \ngc{4388}, namely: \pexrav\ with $\mathcal{R} = 0.5$, a cut-off power law and \compps. We obtained the same improvement in terms of $\chisq$ for all the three models ($\rchisq=5/9$; see Fig. \ref{fig:ratios2110}). Finally, we extrapolated the best-fitting cut-off power law down to 3 keV and compared it with the \xmm\ and \nus\ spectra (Fig. \ref{fig:extrap_2110}). Despite the significant flux variability in the 3--79 keV band, there is no strong evidence of significant spectral variability among the different observations.

\subsubsection{The broad-band fit}
Next, we included the \xmm\ and \nus\ data.
We first fitted the data with a phenomenological model consisting of an absorbed cut-off power law plus a Gaussian \fek\ line. In \xspec, the model reads \textsc{phabs(cutoffpl+zgauss)}. According to the results of \cite{marinucci2015}, the flux of the iron line varies among the \xmm\ and the two \nus\ observations, while the values of the line energy and intrinsic width are consistent with each other within the errors. We thus left the line flux free to vary among the observations, while the intrinsic width was tied.
Keeping all the other parameters tied among the different spectra, we found a good fit ($\rchisq=975/1009$). However, we found an improvement by leaving the absorbing column density free to vary among the spectra, finally obtaining $\rchisq=947/1007$ ($\dchi/\ddof = -28/-2$; the probability of chance improvement is \expo{4}{-7} from an F-test), with $\Gamma \simeq 1.65$ and a $\cut = \aerm{320}{100}{60}$ keV. No significant improvement is found by leaving the photon index free to vary, and the cut-off energy is consistent with the value reported above. 
The best-fitting parameters are reported in Table \ref{tab:fits_2110}, while the contour plots of the cut-off energy and photon index are shown in Fig. \ref{fig:cont_cutoffpl}.
To check for the presence of a reflection component, we replaced the cut-off power law with \pexrav, leaving the reflection fraction $\refl$ free and tied among the observations. The fit is not improved
and we obtain a stringent upper limit $\refl < 0.02$. 

Then, we replaced the cut-off power law with \compps, assuming a spherical geometry and a seed photon temperature of 100 eV.
We obtained a good fit ($\rchisq=971/1007$), with $\kte = \aerm{75}{20}{15}$ keV and $y = \aerm{1.20}{0.01}{0.02}$, corresponding to an optical depth $\tau=\serm{2.1}{0.5}$. The best-fitting parameters are reported in Table \ref{tab:fits_2110}, while the $\kte - y$ contour plots are shown in Fig. \ref{fig:cont_compps}.
\begin{table*}
	\begin{center}
		\caption{ \label{tab:fits_2110} Best-fitting parameters for NGC 2110. (t) denotes a parameters tied between the spectra.} 
		\begin{tabular}{ l c c c c} 
			\hline  \hline 
			& all obs. & XMM & NUS1 & NUS2 \\
			\hline
			\multicolumn{5}{c}{\textsc{cutoffpl} model }\\
			\hline
			\noalign{\smallskip}
			$\Gamma$ & \ser{1.65}{0.02}&(t)&(t)&(t) \\
			$\cut$ (keV)& \aer{320}{100}{60}&(t)&(t)&(t) \\
			$N_{\textsc{pow}}$ (\tento{-2}) & \ser{2.27}{0.09}&(t)&(t)&(t)  \\     
			\noalign{\smallskip}
			$K_{\textrm{IBIS--pn}}$  && \ser{0.34}{0.01} &-&-        \\
			$K_{\textrm{IBIS--NusA}}$  &&-  &\ser{2.63}{0.06}&   \ser{1.79}{0.05}        \\  
			$K_{\textrm{IBIS--NusB}}$  && - &\ser{2.71}{0.06}&   \ser{1.81}{0.05}        \\   
			\noalign{\smallskip}
			$\nh$ (\tento{22} \sqcm) & & \ser{4.4}{0.2}&\ser{3.5}{0.3}&\ser{3.6}{0.3}\\        
			\noalign{\smallskip}
			$E_{\textsc{ga}}$ (keV) & \ser{6.42}{0.01}&(t)&(t)&(t) \\
			$\sigma_{\textsc{ga}}$ (eV) & \ser{40}{20}&(t)&(t)&(t) \\
			$N_{\textsc{ga}}$ (\tento{-4}) && \ser{1.5}{0.2}  & \ser{0.23}{0.11} & \ser{0.72}{0.16} \\
			EW$_{\textsc{ga}}$ (eV) && \aer{140}{10}{20}  & \ser{20}{10}  & \aer{70}{10}{20}\\
			\noalign{\smallskip}
			$\rchisq$ &947/1007&&&\\
			\hline  			
			\multicolumn{5}{c}{\textsc{compps} model }\\
			\hline        
			\noalign{\smallskip}
			$\kte$ (keV) & \aer{75}{20}{15}&(t)&(t)&(t) \\[0.5ex]
			$y$ & \aer{1.20}{0.01}{0.02}&(t)&(t)&(t) \\[0.5ex]
			$N_{\textsc{compps}}$ (\tento{4})  &\aer{5.6}{0.4}{0.3} &(t)&(t)&(t) \\  
			\noalign{\smallskip} 
			$\rchisq$ &971/1007&&&\\  		
			\hline	
		\end{tabular}
	\end{center}
\end{table*}

\begin{figure} 
	\includegraphics[width=\columnwidth]{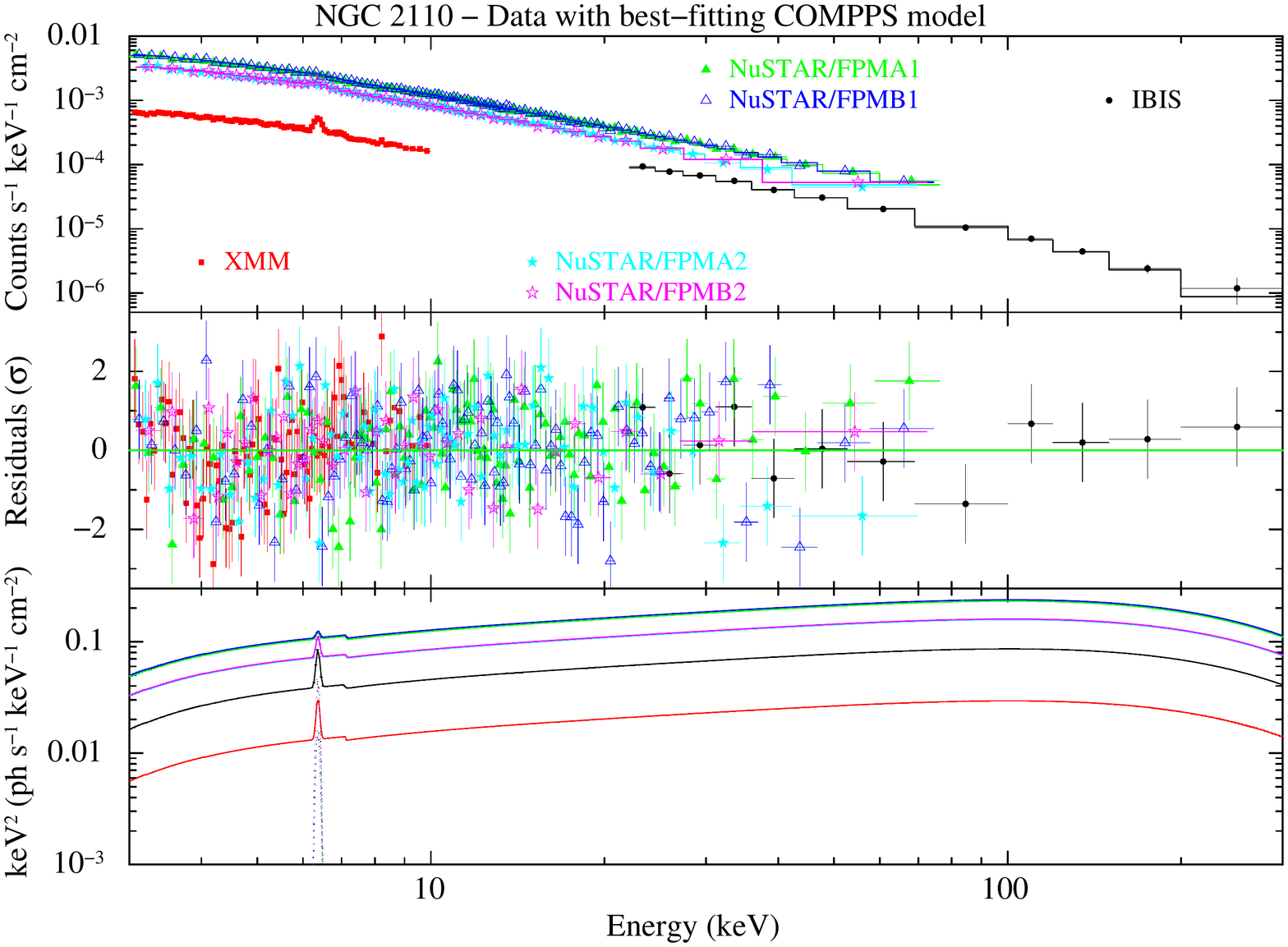}
	\caption{\label{fig:plot_2110} Upper panel: spectra of NGC 2110 with best-fitting \compps\ model. Second panel: residuals, plotted as $\Delta \chi = $ (data-model)/error. Third panel: best-fitting model $E^2 f(E)$. The data were binned for plotting purposes.}
\end{figure}

\section{Discussion and conclusions}\label{sec:discussion}
We presented the hard X-ray spectra of the two X-ray bright Seyferts \ngc{4388} and \ngc{2110}, from \integral's IBIS data taken from 2003 to early 2015.
From a joint fit with \xmm\ and \nus\ data, we find that the spectrum up to 300 keV is in both cases well described by an absorbed power law with a well-constrained high-energy cut-off. In agreement with past observations, we find no evidence of a Compton hump. The primary continuum is generally consistent with being constant in spectral slope among the time-averaged IBIS spectrum and the \xmm\ and \nus\ observations, with some absorption variability seen in \ngc{2110}. This is consistent with the lack of clear evidence of strong hard X-ray spectral variability in these objects \citep{soldi2014}. The exception is the 2011 \xmm\ observation of \ngc{4388}, which has a flatter and less obscured spectrum (see Appendix \ref{appendix}). On the other hand, the cross-normalisation constants indicate a significant flux variability in the \xmm\ and \nus\ energy bands, up to a factor of approximately eight in \ngc{2110}.

Fitting the data with a phenomenological cut-off power law, we obtain photon indices of \aer{1.60}{0.07}{0.06} for \ngc{4388} and of \ser{1.65}{0.02} for \ngc{2110}, both perfectly consistent with the values previously reported from \nus\ data alone \citep{kamraj2017,marinucci2015}. 
As for the cut-off energies, we obtain \aer{200}{75}{40} keV for \ngc{4388} and \aer{320}{100}{60} keV for \ngc{2110}; the latter measurement is consistent with the lower limit of $\sim 200$ keV previously reported from \suz\ and \nus\ data \citep{rivers2014,marinucci2015}. Both values are also consistent within the errors with those reported by \cite{ricci2017}. In the case of \ngc{4388}, the measurement is consistent with previous \integral\ results \citep{fedorova2011,molina2013}. However, the IBIS spectrum analysed here shows a clear and unambiguous high-energy turnover, not related to a putative reflection hump which could affect the results \citep{molina2013}.
The cut-off energies of the two sources are relatively high compared with both the \integral\ high-energy cut-off distribution reported by \cite{malizia2014}, which has a mean of 128 keV with a spread of 46 keV, and with 
the values recently measured with \nus\ \citep{tortosa2018,lanzuisi2019}. 
Among the 19 bright sources observed by \nus\ and discussed by \cite{tortosa2018},
only two have a cut-off $>200$ keV and a coronal temperature $>70$ keV 
with a percent error $\lesssim 50\%$.

From a fit with a thermal Comptonisation model and assuming a spherical geometry, we obtain in both cases a coronal temperature of 70-80 keV and an optical depth close to two. In the case of \ngc{4388}, the temperature of \aer{80}{40}{20} keV is consistent within the errors with the upper limit of 70 keV reported by \cite{lubinski2016}, who use the same model (i.e. \compps\ with spherical geometry). In the case of \ngc{2110}, instead, the temperature of \aer{75}{20}{15} keV is much lower than the value reported by \cite{lubinski2016}, who find a lower limit of 160 keV. However, \cite{lubinski2016} also report a relatively large reflection fraction of $\sim 0.6$, which is not consistent with the \nus\ and \suz\ constraints \citep[as already noted by][Sect. 5.1.3]{lubinski2016}.
We verified that imposing $\refl = 0.6$ we obtain a temperature of $\sim 340$ keV, but with a worse fit ($\Delta \chisq = +12$; F-test probability of \expo{4.4}{-4}).
\cite{lubinski2016} also note that the discrepancy could be due to the 
use of \integral\ data up to 2010 only, 
and that adding more recent \integral\ data result in a lower value of $\refl$.
Our results are roughly consistent with the temperature being a factor of two or three smaller than the cut-off energy, depending on the optical depth \citep{pop5548,poptestingcompt2001}. The difference is actually by a factor of four in \ngc{2110}, but since the high-energy turnover of a Comptonisation spectrum is much sharper than an exponential cut-off \citep[e.g.][]{stern1995,zdziarski2003,nied2019}, this difference should not be overinterpreted \citep{poptestingcompt2001}. 

In both sources, the \fek\ emission line is not accompanied by a significant Compton reflection hump, suggesting that the line originates from Compton-thin material, such as the broad-line region \citep{kamraj2017,marinucci2015} or a Compton-thin torus \citep{borus}. The lack of a Compton hump due to reflection off the accretion disc or an optically thick torus could also indicate that the X-ray corona does not effectively illuminate this surrounding matter. This could be a geometrical effect, if the disc and the torus subtend a small solid angle from the corona. Alternatively, the X-ray coronal emission could be anisotropic \citep{beckmann2004,kamraj2017}. For example, if the X-ray corona is outflowing at relativistic velocities, its emission would be beamed away from the disc, possibly producing a weaker reflection component \citep[][]{malzac2001}.

The 2--10 keV luminosity extrapolated from the IBIS time-averaged spectrum is \serexp{1.3}{0.1}{43} \lumcgs\ for \ngc{4388} and \serexp{1.33}{0.03}{43} \lumcgs for \ngc{2110}, respectively. Using the bolometric correction of \cite{marconi2004}, we estimate a bolometric luminosity of $ \sim \expom{2.5}{44}$ \lumcgs\ for both sources. Given the different black hole mass, this yields a different accretion rate in Eddington units, of $\sim 0.23$ for \ngc{4388} and $\sim 0.01$ for \ngc{2110}. The latter is in agreement with the estimate reported by \citep[][]{marinucci2015}, with the caveat that the estimate of black hole mass in \ngc{2110} has a large uncertainty.

The X-ray corona of AGNs can produce radio synchrotron emission \citep{raginski&laor2016}, as recently observed in the two bright Seyferts IC 4329A and NGC 985 \citep{inoue&doi2018}. This could explain the radio emission seen in radio-quiet Seyferts, whose origin is under debate \citep[][and references therein]{panessa2019}.
In general, a coronal origin for the radio emission is suggested by the observed relation between the radio luminosity at 5 GHz $L_R$ and the X-ray (0.2--20 keV) luminosity $L_X$, with $L_R/L_X=\tentom{-5}$ like in coronally active stars \citep{laor&behar2008}. Another hint is the presence of a mm-wave excess with respect to the spectral slope extrapolated from low frequencies \citep{behar2015,behar2018}. 
\ngc{4388} has a complex radio structure \citep{falcke1998,mundell2000}, with a faint core having a flux density of 1.3 mJy at 1.6 GHz and $<0.55$ mJy at 5 GHz \citep{gp2009}. This source exhibits a mm excess \citep{behar2018}, but it also has a $L_R/L_X$ ratio $< \tentom{-7}$ \citep[from the upper limit to the 5-GHz luminosity reported by][]{gp2009}.
\ngc{2110} shows an S-shaped jet-counterjet radio structure, with a relatively bright core having a flux density of 100 mJy at 5 GHz \citep{ulv1983,mundell2000}. \cite{beckmann2010} argue that this object is borderline between a radio-quiet Seyfert and a radio-loud galaxy with a \cite{fr1974} type I morphology.
No mm excess is found in \ngc{2110}, whose radio and mm emission is consistent with originating from the jet \citep{behar2018}. 
We conclude that the radio emission is unlikely to be of coronal origin in the two sources discussed here. 
According to our results, \ngc{4388} and \ngc{2110} have a corona with similar physical parameters and producing the same X-ray luminosity. However, they seem to have different accretion/ejection parameters: while \ngc{4388} has a high Eddington ratio and a weak radio core, \ngc{2110} has a relatively low Eddington ratio and a bright radio core. This is consistent with the inverse relation between radio loudness and Eddington ratio, found both in radio-quiet and radio-loud objects \citep{ho2002,sikora2007,panessa2007}. The physical origin of this anticorrelation is a matter of debate \citep[][]{broderick&fender2011,beckmann&schrader}.

Finally, we put \ngc{4388} and \ngc{2110} in the compactness-temperature ($\ell - \tetae$) diagram \citep{fabian2015} by calculating the dimensionless coronal temperature $\tetae = \kte/\elmass c^2$ and the compactness parameter $\ell = L \sigma_\textrm{T} / R \elmass c^3$, where $L$ is the luminosity and $R$ is the radius of the corona. Following \cite{fabian2015}, we used the luminosity of the primary continuum extrapolated to the 0.1--200 keV band. For both sources we assumed the flux level of the IBIS spectra. Denoting by $R_{10}$ the coronal size in units of 10 gravitational radii, we obtained 
$\ell \simeq 130 \, (R_{10})^{-1}$ 
and $\tetae = \aerm{0.16}{0.08}{0.04}$ for \ngc{4388}, while for \ngc{2110} we obtained 
$\ell \simeq 5 \, (R_{10})^{-1}$ 
and $\tetae = \aerm{0.15}{0.04}{0.03}$. Then, assuming a radius of 10 gravitational radii, both sources would lie below the pair runaway line, that is a maximum-luminosity curve delimiting a forbidden region in which pair production would exceed annihilation \citep{fabian2015}. Our results are thus in agreement with a scenario in which the so-called pair thermostat \citep[e.g.][]{svensson1984,zdziarski1985} controls the coronal temperature in AGNs \citep{fabian2015}, possibly with a contribution from non-thermal particles \citep{fabian2017}.
The compactness estimated for \ngc{2110} is not very high, but given the uncertainty on the black hole mass (and on the coronal size), it could easily be $>10$.
Interestingly, for a pair-dominated corona, the spectral shape is almost constant unless the luminosity variations are strong \citep[e.g. by a factor of 20: see][]{hmg1997}. 

We note that the detection of a high-energy roll-over in the two sources discussed above is enabled by a broad-band energy coverage, extending up to 300 keV thanks to the unique capabilities of IBIS.
These results demonstrate that spectral information above 100 keV is crucial to constrain the properties of the hot corona in AGNs, especially in the high-temperature regime.
Extending the census of AGN coronal parameters is in turn essential to test the pair thermostat scenario and to search for correlations between the different coronal parameters, that may carry significant information on the disc/corona system \citep{tortosa2018,middei2019}.   
Future high-energy missions like \textit{HEX-P} \citep{hexp} and \textit{ASTENA} \citep{fuschino2018} will be key to provide single-epoch measurements of the high-energy cut-off. 

\begin{acknowledgements}
	We thank the referee for comments that improved the manuscript.
We acknowledge financial support from ASI and INAF under INTEGRAL `accordo ASI/INAF 2013-025-R1'. 
\end{acknowledgements}

   \bibliographystyle{aa} 
   \bibliography{mybib.bib} 
%

\appendix
\section{The 2011 \xmm\ spectrum of \ngc{4388}}\label{appendix}
\begin{figure} 
	\includegraphics[width=\columnwidth]{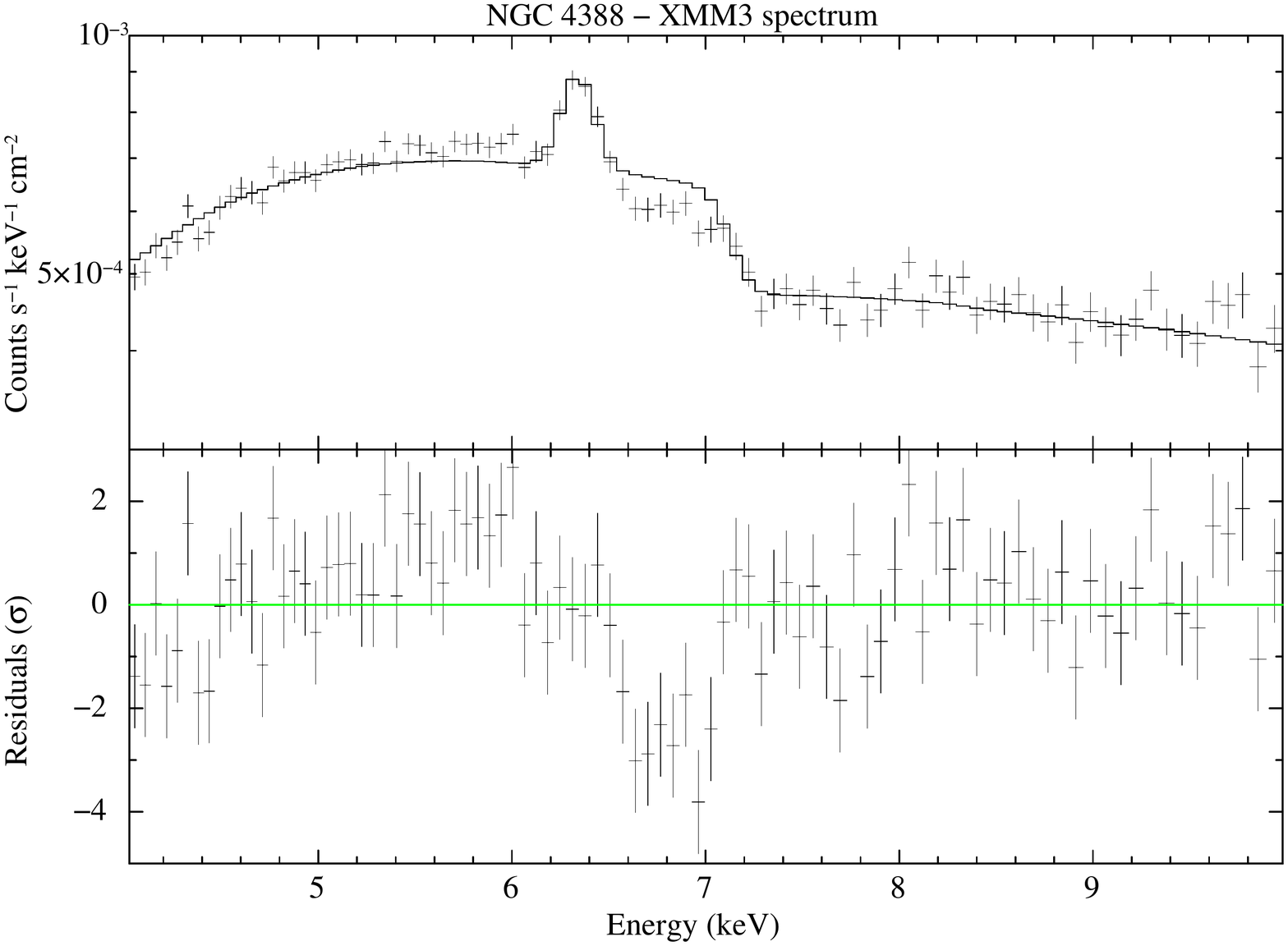}
	\caption{\label{fig:xmm3} Upper panel: 2011 \xmm\ spectrum of \ngc{4388} fitted with a an absorbed power law plus a narrow Gaussian line. Lower panel: residuals as $\Delta \chi$.}
\end{figure}

To analyse the 4--10 keV spectrum of the XMM3 obs. (see Table \ref{tab:log}), we first assumed the best-fitting model described in Sect. \ref{sec:analysis}, i.e. a cut-off power law plus a narrow Gaussian line modified by partial covering absorption. We fixed the cut-off energy at 200 keV, since we used only low-energy \xmm\ data. We obtained a poor fit with $\rchisq=153/85$ and strong, negative residuals in the 6.5-7.0 keV band (see Fig. \ref{fig:xmm3}). We thus added one narrow Gaussian line in absorption, finding an improved but still unacceptable fit ($\rchisq=116/83$). Then, we added a further absorption line, obtaining a good fit with $\rchisq=71/81$. The best-fitting parameters are reported in Table \ref{tab:fit_xmm3}. The photon index is \aer{1.50}{0.02}{0.08}, that is to say flatter than the value of \aer{1.60}{0.07}{0.06} reported in Sect. \ref{sec:analysis}. The absorbing column density is also lower, amounting to $\sim \expom{2.4}{23}$ \sqcm\ instead of $\sim \expom{3.8}{23}$ \sqcm. The two absorption lines are found at rest-frame energies of \aer{6.72}{0.02}{0.03} keV and  \aer{6.99}{0.04}{0.02} keV, and can be identified as the \kalfa\ lines of \ione{Fe}{xxv} at 6.7 keV and of \ione{Fe}{xxvi} at 6.966 keV, respectively.
Ionized iron absorption lines have been detected in a number of objects. In many cases the line energies are blueshifted, implying an origin from a fast outflow \citep[e.g.][]{tombesi2010}. However, absorption from `static' or low-velocity ionised material has also been observed in some sources \citep[][and references therein]{bianchi2005}. For example, \cite{reeves2004} detected a variable absorption line from highly ionised iron in an \xmm\ spectrum of the Seyfert 1 \ngc{3783}. This line was found to be stronger at higher continuum flux levels, and it was ascribed to a component of the warm absorber present in that source \citep{reeves2004}. 
A proper test of this hypothesis is 
difficult in an obscured object like \ngc{4388}, 
and is out of the scope of the present work. However, following \cite{reeves2004}, we can tentatively ascribe the emergence of ionised iron absorption lines in the high-flux 2011 spectrum to a change in the ionisation level of the circumnuclear absorbing material due to an increase of the illuminating flux.
 
\begin{table}
	\begin{center}
		\caption{ \label{tab:fit_xmm3} Best-fitting parameters for the XMM3 obs. of \ngc{4388}. The model includes an absorbed power law plus three narrow Gaussian lines (line 1 and 2 in absorption, line 3 in emission).} 
		\begin{tabular}{ l c } 
			\hline  \hline 
			$\Gamma$ & \aer{1.50}{0.02}{0.08} \\
			$N_{\textsc{pow}}$ (\tento{-2}) & \aer{1.8}{0.1}{0.3}\\
			\noalign{\smallskip}
			$\nh$ (\tento{23} \sqcm) & \aer{2.4}{0.1}{0.4} \\        
			$\cf$   & \linf{0.95} \\
			\noalign{\smallskip}	
			$E_1$ (keV) & \aer{6.72}{0.02}{0.03} \\	
			$N_1$ (\tento{-4}) & \ser{-3.2}{0.7} \\
			EW$_1$ (eV) & \ser{-30}{10} \\
			\noalign{\smallskip}	
			$E_2$ (keV) & \aer{6.99}{0.04}{0.02} \\
			$N_2$ (\tento{-4}) & \ser{-3.2}{0.7} \\
			EW$_2$ (eV) & \ser{-40}{10} \\
			\noalign{\smallskip}	
			$E_3$ (keV) & \aer{6.40}{0.01}{0.02} \\	
			$N_3$ (\tento{-4}) & \ser{5.7}{0.8} \\	
			EW$_3$ (eV) & \ser{50}{10} \\
			\hline				
		\end{tabular}
	\end{center}
\end{table}

\end{document}